\documentstyle [12pt,epsfig]{article}

\def\ref{\bibitem{}}
\begin{document}
\vspace*{5mm}
\begin{center}
{\Large \bf X-ray Shots of Cyg X-1
 \footnote{To appear in {\sl The Astrophysical Journal}}}
 
\vspace{5mm}
Y. X. Feng~~~T. P. Li\\  
{\sf High Energy Astrophysics Lab., Institute of High Energy Physics, Beijing}\\

\vspace{3mm}
L. Chen\\
{\sf Department of Astronomy, Beijing Normal University, Beijing}
\end{center}

\begin{abstract}
X-ray shots  of Cyg X-1 in different energy bands 
and spectral states have been studied with PCA/{\sl RXTE} observations.
The detailed shot structure is obtained by superposing many shots with 
one millisecond time bin through aligning their peaks with an improved 
algorithm. In general, the shots are composed of a slow rise and fast
decay. The shot structures in the different states are different.  
The duration of shot in the high state is shorter than that in the low
and transition states.
The shot profile in the high energy band is more asymmetric
and narrower than that in the low energy band.  
The average hardness of shot is lower than that of steady
emission in the transition and low states but higher than that in the high state.
The time lags between the shots in higher and lower energy bands
have been found in the different states. In transition states, the time lag
is the largest among the different states of Cyg X-1, and it is the smallest
in the low state. 
The implications of the observed shot features
for shot models are discussed.\\
{\sl Subject headings:} accretion: accretion disk - stars: individual 
(Cygnus X-1) - X-rays: stars
\end{abstract}

\section{Introduction}
The black hole binary Cyg X-1 has complex
X-ray emission variability at all the  time scales from milliseconds to years
(cf. review by Tanaka \& Lewin 1995, van der Klis 1995). 
On the short time scale of subsecond the hard X-rays from Cyg X-1 show large-amplitude 
chaotic fluctuations. 
The studies of temporal properties of the rapid variability
in the hard x-rays should be a hopeful method to understand the properties of the 
compact object. In the low state (LS), the variability of X-rays could be
described approximately as the superposition of individual shots
(Terrel 1972; Oda 1977; Sutherland et al. 1978; Miyamoto et al. 1988;
Miyamoto et al. 1992). The properties of the rapid fluctuation
or shot had been studied from ACF (Autocorrelation Function) and PDS 
(Power Density Spectrum) based on the supposition of randomly occurring 
shots (Nolan et al. 1981; Lochner et al. 1991). 
Negoro, Miyamoto and Kitamoto (1994) (hereafter NMK) obtained average shot profiles 
from a {\sl Ginga} observation of Cyg X-1 in the low state by superposing many shots 
through aligning their peaks and found that 
the shot profile was approximately represented by a sum of two exponentials with time
constants of about 0.1 and 1 s and that
the average hardness during the shot is lower  than that
of steady emission.

Quite a few models have been proposed as the mechanism for
creating the rapid fluctuations, such as the hot spot model
in the inner part of disk (Lightman \& Eardley 1974; Shibazaki \& Hoshi 1975;
Rothschild et al. 1974; Miyamoto \& Kitamoto 1989), the disturbance
propagation model (Manmoto et al. 1996) and the model of magnetic flare 
(Galeev et al. 1979; Pudritz \& Fahlman 1982; Nolan et al. 1981; Lochner
et al. 1991) .  The rapid hardening of the spectra at the peak 
obtained from the study of average shot features by NMK  
is difficult for the magnetic flare model to explain. 
Recently, the observed phase lags and variability coherence between 
high- and low-energy bands have ruled out the magnetic flare model 
and the models of hot spot, and it seems that the simple propagation model 
could qualitatively represent the 
properties of x-ray fluctuation (Nowak et al. 1997; Vauhgan \& Nowak 1997). 

Up to now, the shot properties have only been studied with observations in the 
low state of Cyg X-1.
The {\sl RXTE} mission (Bradt et al. 1993) provided a unique 
opportunity  for shot property studies in Cyg X-1 with microsecond resolution
observations  when Cyg X-1 experienced transitions from
the low state to the high state, and then from the high state to the low state
( Belloni et al. 1996; Cui et al. 1997a, 1997b, Zhang Cui et al. 1997). In this 
paper, we present our analysis of the structures of X-ray shots in the different states
of Cyg X-1 with an improved method from {\sl RXTE} observations.

\section{Observations and Data Analysis}
\subsection{Observations} 
There are fifteen public observations with {\sl RXTE} from May 22 
to August 15, 1996, which covered transition states (IS) and high 
state (HS) of Cyg X-1 (Belloni et al. 1996; Cui et al. 1997a, 1997b).
The PCA observations were carried out consistently with three modes.
Two of the modes were selected for analyzing shot structure with one
millisecond time bin in this work: one is the Event-mode with sixteen energy bands covering 
$13-60$ keV with the time resolution of 16 $\mu$s; the other is the Single-Bit 
mode including $2-6$ keV and $6-13$ keV energy bands with the time bin of 
125 $\mu$s. The data from two of PCA/{\sl RXTE} observations on 1996
Dec. 17 when Cyg X-1 stayed in the low state, were also used in this
study. In those data, the energy bands of Single-Bit modes of PCA
are $2-5$ keV and $5-13$ keV with the time bin of 125 $\mu$s.

The chosen data have been further binned with a $2^{-10}$ s time resolution.
The collimator response correction and barycentric correction are applied. In the 
study of the average features of shots from a bright source, such as Cyg X-1, 
the background can be negligible. The effect of dead time can be also neglected
because the real count rate is not very high and the time bin is much larger 
than the 10$\mu$s dead time (Zhang 1995).  

\subsection{Analysis Methods}
According to  NMK, shots were selected by the criteria that their
peak count of $1.2-58.4$ keV X-rays should be $2-3$ times larger than that of the
local mean counts obtained in the 32 s interval and should be the maximum
within the neighboring 8 s on both sides. 
The average shot profile with the 15.6 ms time bin in the $1.2-58.4$ keV energy 
band and that in the five subdivision bands ($1.2-7.3, 7.3-14.6, 
14.6-21.9, 21.9-36.5$ and $36.5-58.4$ keV) were obtained separately 
by superposing the count profiles in each band of selected shots through 
aligning their peak bins determined by the total light curve in $1.2-58.4$ keV.  
Since the differences between the times of shot peaks in different energy bands 
were not taken into account and that the shots near which there were other smaller 
shots within 8 s were not excluded in the NMK's procedure, the average shot 
structure may deviate from the truth. 
Li and Fenimore (1996) (hereafter LF) proposed a peak-finding algorithm to analyze 
the $\gamma$-ray burst temporal structure. In this algorithm, a candidate peak 
is selected as the bin having more count ($C_{p}$) than the neighboring bins 
(both sides), 
and then both sides of each candidate peak is searched for the bins with count 
$C_{1}$ (at $t_{1} < t_{p}$) and $C_{2}$ (at $t_{2}>t_{p}$) so that 
the condition
\begin{equation}
C_{p} > C_{1,2}+\alpha\sqrt{C_{p}+C_{1,2}}
\end{equation}
is satisfied. When both $C_{1}$ and $C_{2}$ are found the candidate peak 
is selected as the true peak of a pulse. This algorithm is sensitive to selecting
the shots but the statistical fluctuation of count has more influence
on its result.   

In order to analyze light curves with a time resolution down to 1 ms 
and low counts in each bin, we combined the two methods mentioned above
and made necessary improvements in our work. To suppress the effects of 
statistical fluctuation, the larger time bin of 10 $\times$ $2^{-10}$s 
($\sim 10$ ms) light curves is used initially to find the shot peak according to
the LF method. Then each selected peak bin and its 
neighboring bins (both sides) are divided into thirty bins with time bin 
of $2^{-10}$ s ($\sim 1$ ms).  A shot peak is selected from the thirty bins 
by using NMK method in which the average count is used 
as the count of the steady emission in the interval of 1 s around the peak. 
This selecting process has been  carried out separately in each energy band of three 
subdivisions between 2-60 keV. A shot peak candidate is selected as a true 
peak if it coincidences in all the three bands within 30 ms. 
To reduce the effects of shot overlap, we excluded any shots near which 
other shot was found within one second. Because there are time delays 
between X-ray intensity variations at different energies, the shot peaks 
were determined and aligned to obtain the average shot profiles separately in
each energy band. 

The improved method used in this work, the NMK method and the LF method
have been compared by simulations with a simulated light curve which has a similar 
count rate of steady and the shot peak as observed by PCA/{\sl RXTE} and 
a shot profile of exponential decay. The obtained average shots are 
displayed in Figure 1. It shows that the average shot feature obtained 
by our method is more similar to the given one than that of  the NMK  
and LF methods. The average shot was not significantly changed as the value 
of parameter $\alpha$ in the selection condition (1) varied from 2.0 to 4.0. 
It means that the shots selected under the condition of $\alpha$=2.0 are 
most likely to be rapid flares with time scale of 10ms, which significantly 
exceed the steady emission, rather than statistical fluctuations. 
In the following analysis, $\alpha$ is set to be 2.0. It can be seen 
that the obtained average shots tend to be expanded because of the selecting 
errors of peak bins due to their count fluctuation.  
The peak bin is excluded in our studies because its count has 
a systematic error caused by the criterion of shot peak selecting.

\begin{figure}
\epsfig{figure=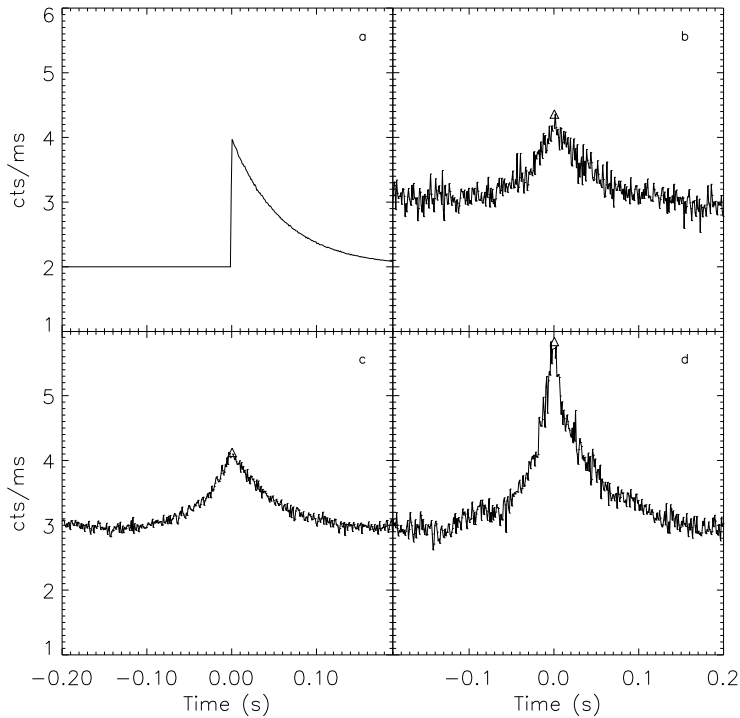,width=10cm,height=10cm,angle=0,
%bbllx=100,bblly=370,bburx=496,bbury=542
}
%\vspace{-3.5cm}
\caption{ Comparison of the present method and other methods. 
(a) Original shot.
(b) The average superposed shot obtained by the method of Negoro et al (1994).
(c) The average superposed shot obtained by the method of Li and Fenimore (1996).
(d) The average superposed shot obtained by the present method.
}
\end{figure}

 We divided the selected shots into two groups with high and low peak counts,
and found that the average shot profiles were very similar between them, as shown 
in Figure 2. It means that the shot profile in one state does not correlate
significantly with the peak counts. A similar result was also obtained by 
NMK from a {\sl Ginga} observation when Cyg X-1 stayed in the 
low state. 
\begin{figure}
\epsfig{figure=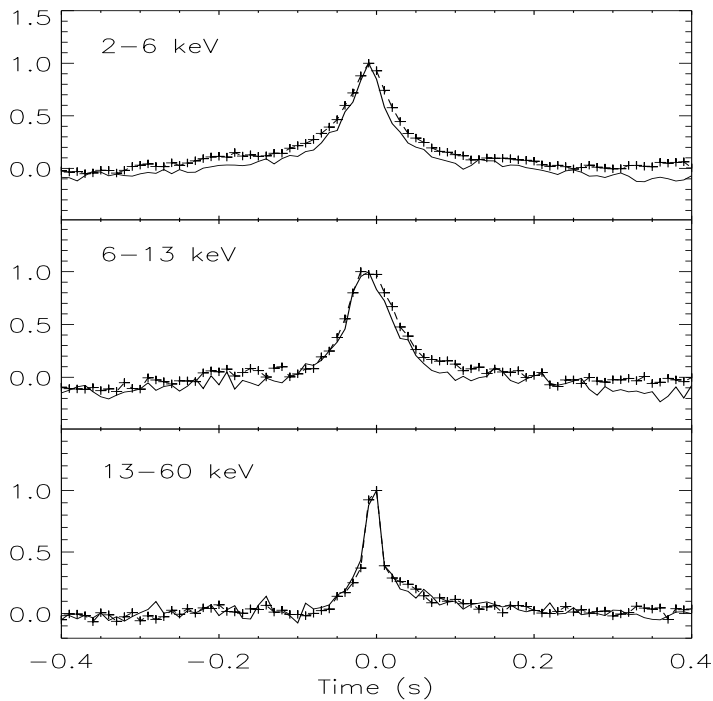,width=10cm,height=10cm,angle=0,
%bbllx=100,bblly=370,bburx=496,bbury=542
}
{\caption{ The average profiles of shots in different energy
bands with different peak intensities. The peak heights are normalized
to be unity. The solid lines represent the average profiles superposed
for the high intensity shots with the peak count 
$C_{p} > C_{b}+4\sigma$,
the crosses for the low intensity shots with  $C_{p} < C_{b}+4\sigma$, 
where $C_{b}$ is the count of steady emission, $\sigma$=$\sqrt{C_{p}+C_{1,2}}$ .}
}
\end{figure}

We used the full width of half maximum (FWHM) of autocorrelation
to describe the duration of shot. The shots were divided into two groups
with duration being longer or shorter than 130 ms. The average shots 
of the different groups are shown in Figure 3. It can be seen that,
except the duration, the shot features are quite similar between 
the long and short duration groups.  The higher the energy bands are, 
the more similar the two groups are. The difference between the two groups 
can be attributed mainly to the slow variation.
The rapid decay processes are almost the same in the two groups.
The hardness evolution during the shots in the two groups is similar too. 
Therefore, the average shot obtained by superposing all the selected shots 
in our studies can reflect the main characteristics of shots with the
different durations and peak intensities, especially the rapid
variability properties.

\begin{figure}
\epsfig{figure=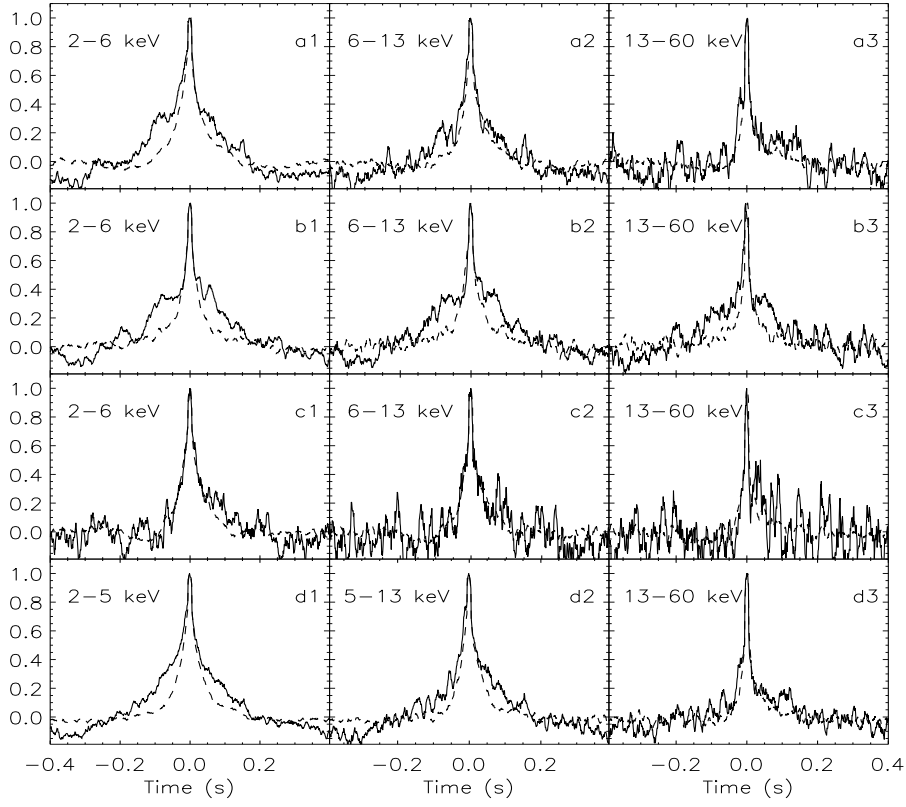,width=13cm,height=11cm,angle=0,%
%bbllx=100,bblly=370,bburx=496,bbury=542
}
\caption{ The superposed shot profiles with the different durations, 
the solid lines correspond to shots with shot durations shorter than
130 ms, and the dash lines correspond to shots with durations longer than 130 ms.
The peak heights are normalized to be unity. 
a1)-a3) for the low to high transition;
b1)-b3) for the high state; c1)-c3) for the high to low transition; d1)-d3) for
the low state.}
\end{figure}

The present method was also applied to the Crab data obtained
with PCA/{\sl RXTE}. The average short time scale ( $\sim$ 10 ms) flare
obtained is similar to the one of the two peak average pulse profile
found by folding with period of 33.4 ms, in duration and rough shape,
as shown in Figure 4. It can be found some smaller flares which
reflect the periodic double-peak pulse structure of Crab. It means that
the flares selected, using our method, are mostly pulses of
Crab. But the double-peak structure found by using our method is not
as clear as by period folding, it is due to most of
periodic pulse selected by us has only one peak of its double-peak pulse
which has significantly positive fluctuation to be selected by our criterion.
It shows that our method can select the real rapid flares and obtain the rough 
average shape of flares which reflect some properties,such as main flare's rise of 
the flares in a real light
curve obtained by PCA/{\sl RXTE} in time scale of $\sim$ 10 ms.

\begin{figure}
\epsfig{figure=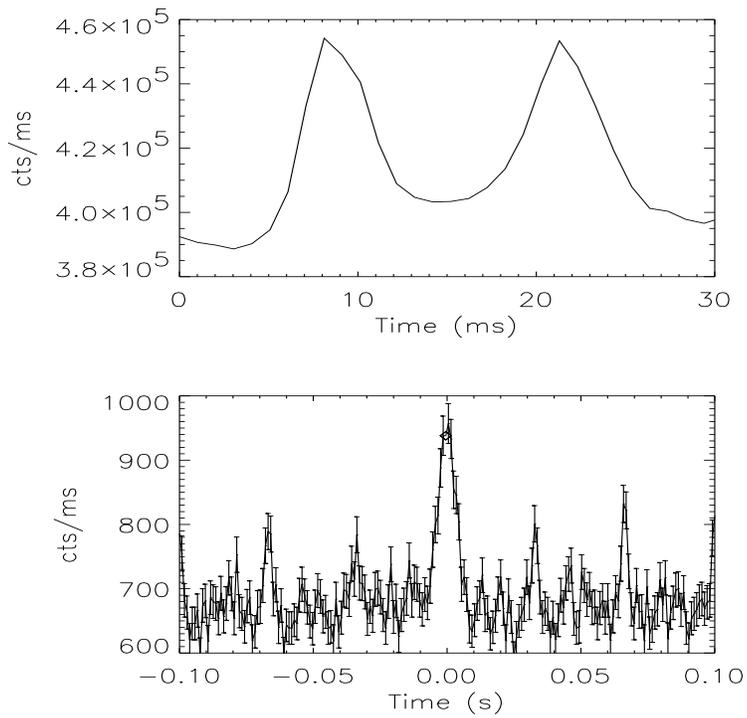,width=10cm,height=10cm,angle=0,%
%bbllx=100,bblly=370,bburx=496,bbury=542
}
\caption{ The comparison between the superposed flare profile and the 
pulse profile in Crab data
with PCA/{\sl RXTE}. a) Pulse profile obtained by period folding. 
b) Superposed flare profile
obtained by our method.}
\end{figure}

\section{Results}
\subsection{Temporal Variability}
 The average profiles of X-ray shots have been obtained from PCA/{\sl RXTE}
observations in the low-to-high transition, high, high-to-low transition and the low  
states of Cyg X-1 by our improved superposing procedure described in
the last section and shown in Figure 5. 
\begin{figure}
\epsfig{figure=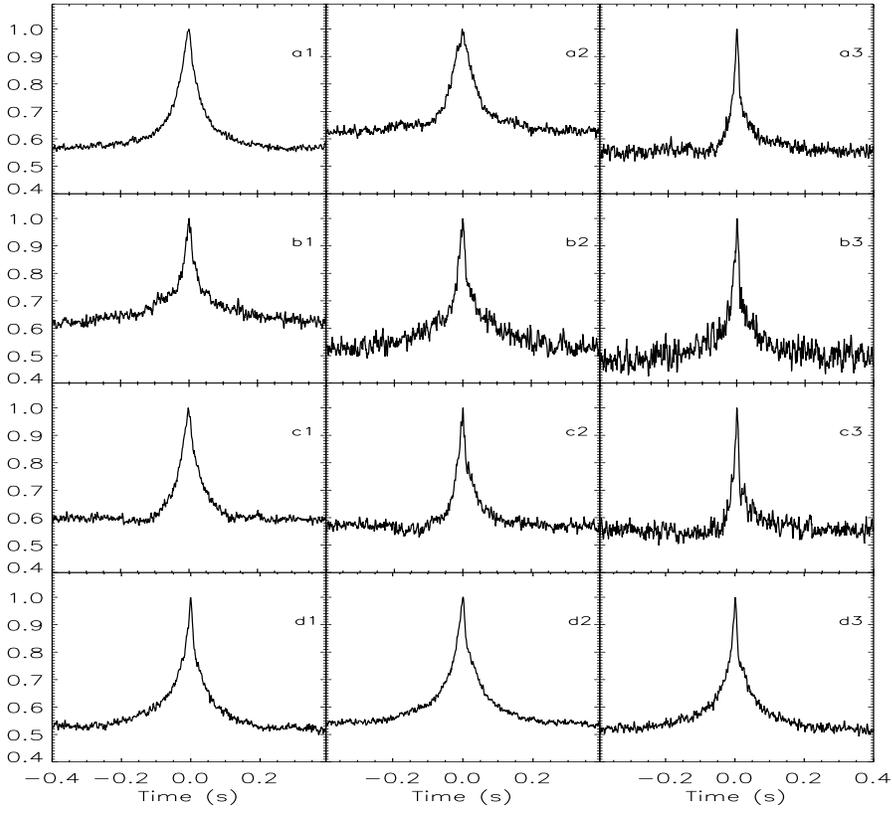,width=13cm,height=11cm,angle=0,%
%bbllx=100,bblly=370,bburx=496,bbury=542
}
\caption{ The superposed shot profiles. The peak heights are normalized to be unity.
The profiles a1-a3 are obtained by superposing 537 shots of the low to high state 
transition in the energy bands of 2-6, 6-13 and 13-60 keV respectively;
b1-b3 by 438 shots of the high state in the 2-6, 6-13 and 13-60 keV bands respectively;
c1-c3 by 268 shots of the high to low state transition in the 2-6, 6-13 and 13-60 keV
bands respectively;
d1-d3 by 564 shots of the low state in the 2-5, 5-13 and 13-60 keV bands respectively.}
\end{figure}
In the transition states, the front of shot could be simply represented by 
a single exponential. However, in the high and low states, a sum of two 
exponentials which correspond to the rapid and slow variability respectively, 
is necessary. The formula for fitting is
\begin{equation}
I(t)= 
\left\{
            \begin{array}{ll}
$$ A_{-} e^{-\frac{t}{\tau_{-}}} + B_{-} e^{-\frac{t}{\eta_{-}}} + C_{-} $$ & ( t < 0 )  \\
$$ A_{+} e^{-\frac{t}{\tau_{+}}} + B_{+} e^{-\frac{t}{\eta_{+}}} + C_{+} $$ & ( t > 0 ) 
\end{array}   
 \right.
\end{equation}
where  $\tau_{-}$ and $\tau_{+}$ correspond to the rapid process of shot;  
$\eta_{-}$ and $\eta_{+}$ correspond to a slow process of shot; $t$ is the time 
relative to the shot peak. The best-fit parameters for the shots, 
within 0.45 s or 450 bins with 1 ms time bin on both sides of shot peak, 
are summarized in Table 1. It shows that the rise of the average shot is 
slower than its decay. And the difference between the slopes of rise and decay 
in the high energy band is more significant 
than that in the low energy band.
\begin{figure}
\epsfig{figure=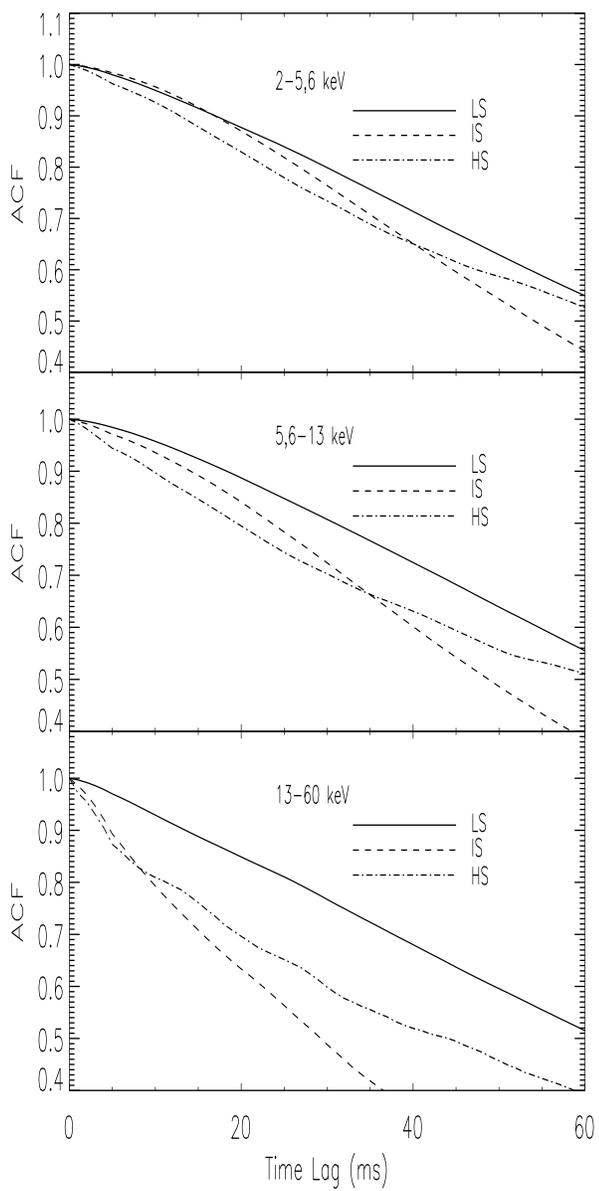,width=8cm,height=16cm,angle=0,%
%bbllx=100,bblly=370,bburx=496,bbury=542
}
\caption{ Autocorrelations of the average shot profiles in the different states.}
\end{figure}
\begin{figure}
\epsfig{figure=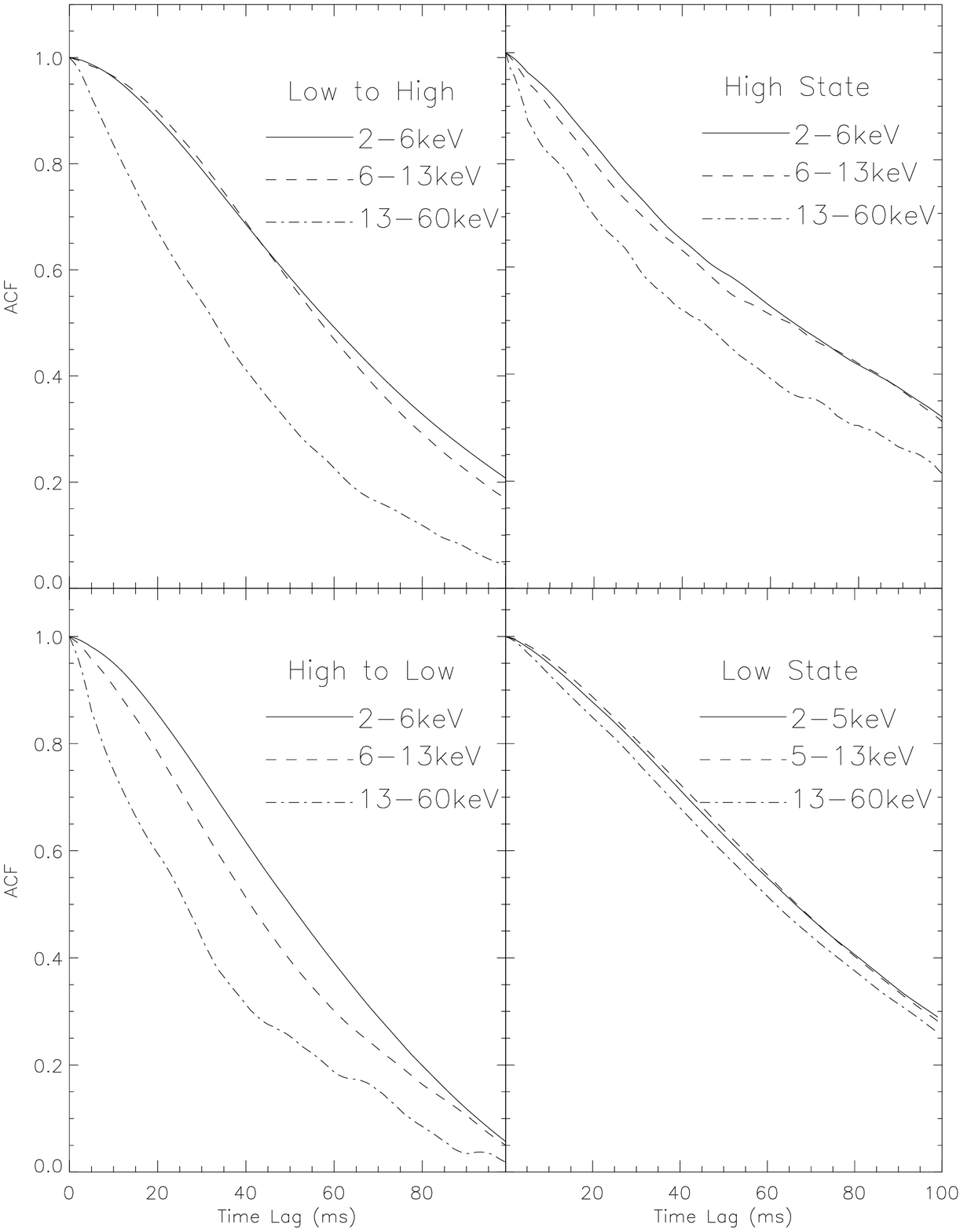,width=13cm,height=9cm,angle=0,%
%bbllx=100,bblly=370,bburx=496,bbury=542
}
\caption{ Autocorrelations of average shot profiles in the different energy bands.}
\end{figure}

The comparison of shot features in the different states and energy bands
are shown in Figures 6 and 7. It can be found that the profiles of shots
depend upon state and energy. The shot duration is the shortest in the high
state and the longest in the low state. The shot in the high energy band is
narrower than that in the lower energy bands. We also found that the time
constant for the rapid decay in the high energy bands is always about
6 ms in the three states.

The distributions of the time lags between the shots of the high and low energy 
bands in the different states are shown in Figure 8. The time lag is determined by
the maximum of the cross-correlation function between the shot profiles in the two
energy bands.  The average lags are listed
in Table 2, where the errors are estimated by the bootstrap
method (Diaconis $\&$ Efron, 1983). It can be seen from Figure 8 and Table 2
that there exists, on the average, a delay of high-energy photons relative 
to low-energy ones in the X-ray shots of Cyg X-1 in each state,
and in the transition sates the distributions of time lags 
are most asymmetric and  the average lags are more larger
 than that in the high and low states.

\begin{figure}
\epsfig{figure=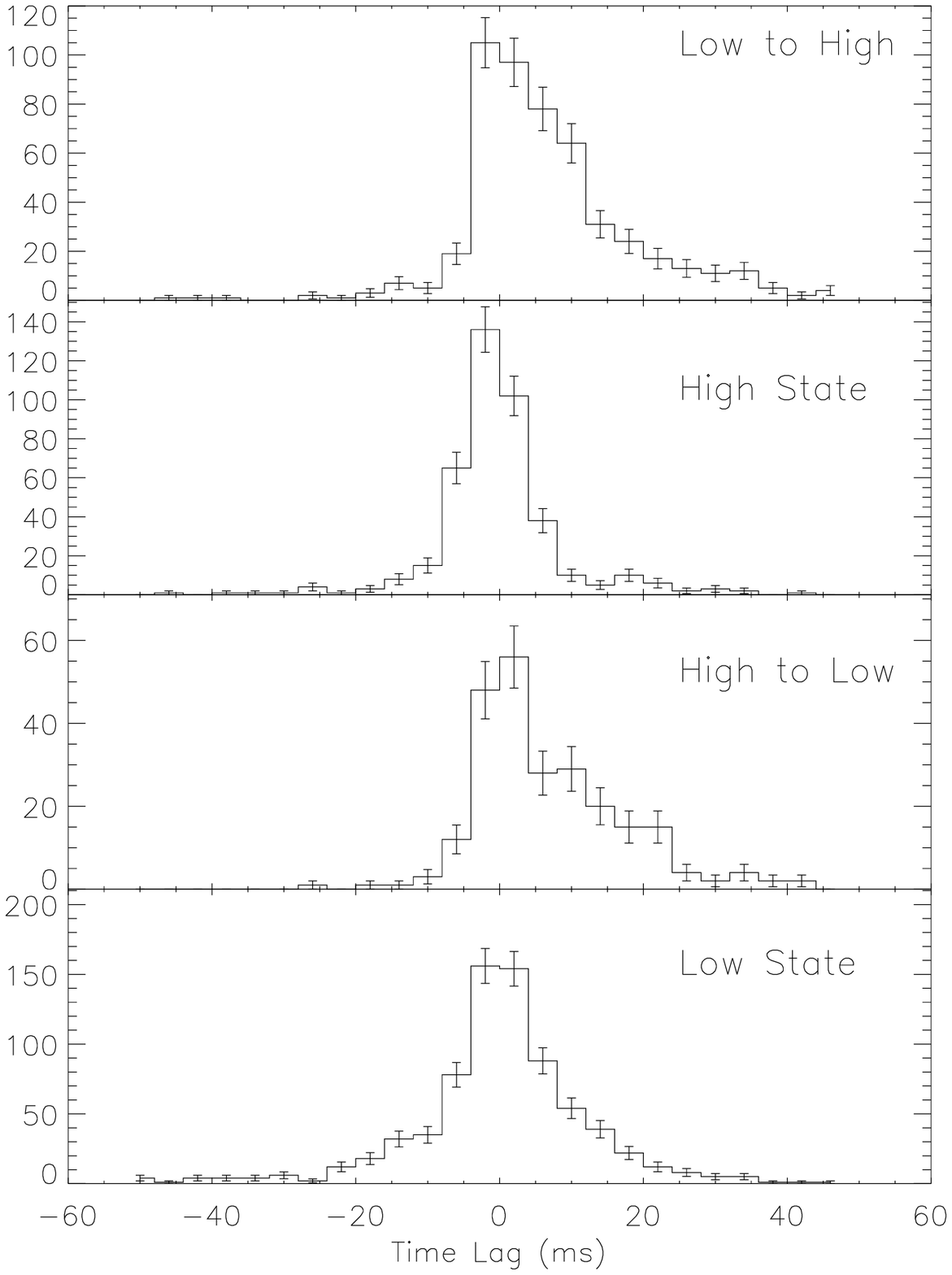,width=13cm,height=11cm,angle=0,%
%bbllx=100,bblly=370,bburx=496,bbury=542
}
\caption{The distributions of time lag between the shots in the
low and high energy bands in different states.}
\end{figure}

\subsection{Spectral Properties}

Figure 9 shows the hardness evolution of average shots in the different states.
The zero points of time in Fig.9 correspond to the shot peaks.
To study the  hardness during a shot, the shot peaks in each 
energy bands have been taken as the bins which are the maximum count bins
in the summed up shots of every energy bands in 2-60 keV. And the time
bin of light curves are set to be 10 $\times$ $2^{-10}$ s to suppress statistical
fluctuation. 
\begin{figure}
\epsfig{figure=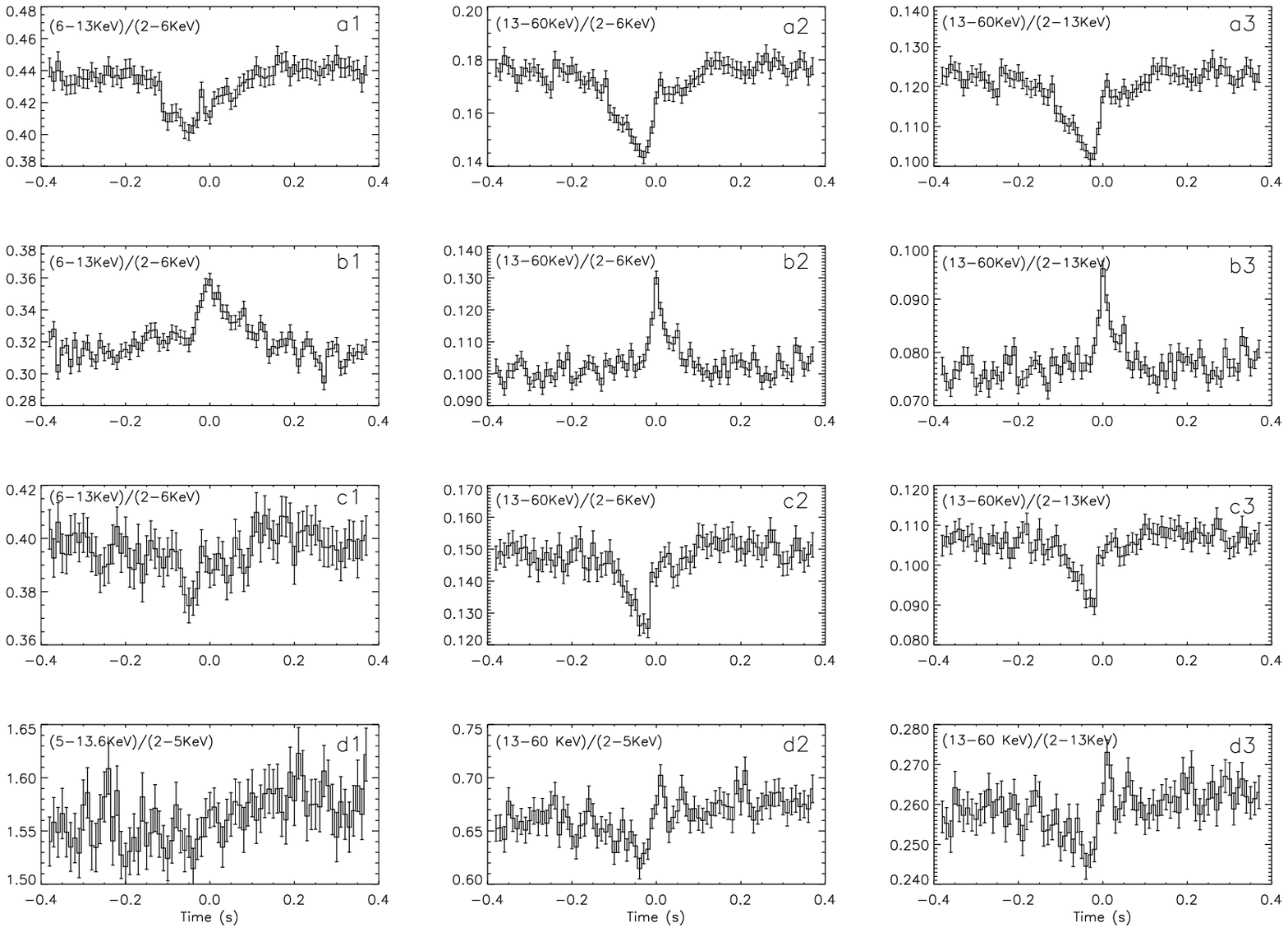,width=13cm,height=11cm,angle=0,%
%bbllx=100,bblly=370,bburx=496,bbury=542
}
\caption{ The hardness ratio profiles of average shots in the different states. 
a1-a3 for the low-to-high transition;
b1-b3 for the high state; c1-c3 for the high-to-low transition; 
d1-d3 for the Low state.}
\end{figure} 
From Fig.9 one can see that in the high state the hardness during a shot is higher than that 
of the steady emission around the shot . 
However, in the low and transition states, the hardness is lower than that of the 
steady emission and the hardness
reduces significantly about 100 ms before the shot peak and rises rapidly to the  
maximum near the peak. This result is similar to that obtained by NMK.
The hardness ratios for the net shots 
after subtracting the steady emission in the different states are  
shown in Figure 10. It shows that after a drop the hardness rises gradually
and reaches to a flat top during the  shot in the low or transition states. 

\begin{figure}
\epsfig{figure=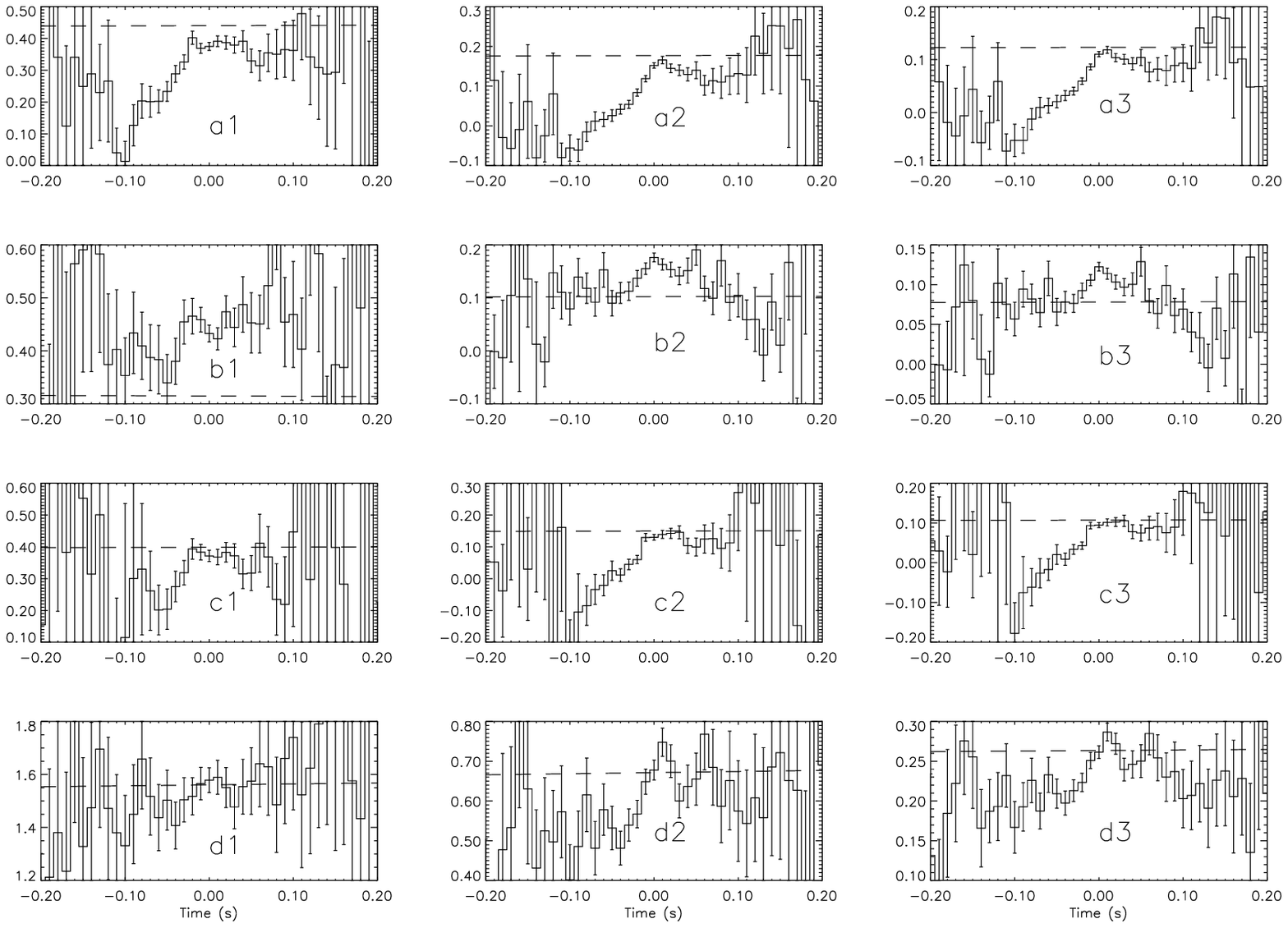,width=13cm,height=11cm,angle=0,%
%bbllx=100,bblly=370,bburx=496,bbury=542
}
\caption{ The hardness ratio profiles of average net-shots in the different states.
The dashed lines correspond to the average off-shot hardness.
a1)-a3) for the low-to-high transition; b1)-b3) for the high state;
c1)-c3) for the high-to-low transition; d1)-d3) for the low state.}
\end{figure}
We take the region from $-0.4$ s to $-0.2$ s and from 0.2 s to 
0.4 s from the shot peak as the off-shot region and the interval between 
$-0.2$ s and 0.2 s around the peak as the shot region. 
The net-shot spectra can be obtained by subtracting the off-shot spectra 
from the shot spectra. The average net-shot
spectra in a state can be obtained by merging all the net-shot spectra
in the state and the average off-shot spectra obtained by
merging all the off-shot spectra. Figure 11 shows the obtained net-shot 
and off-shot spectra in each state separately.
\begin{figure}
\epsfig{figure=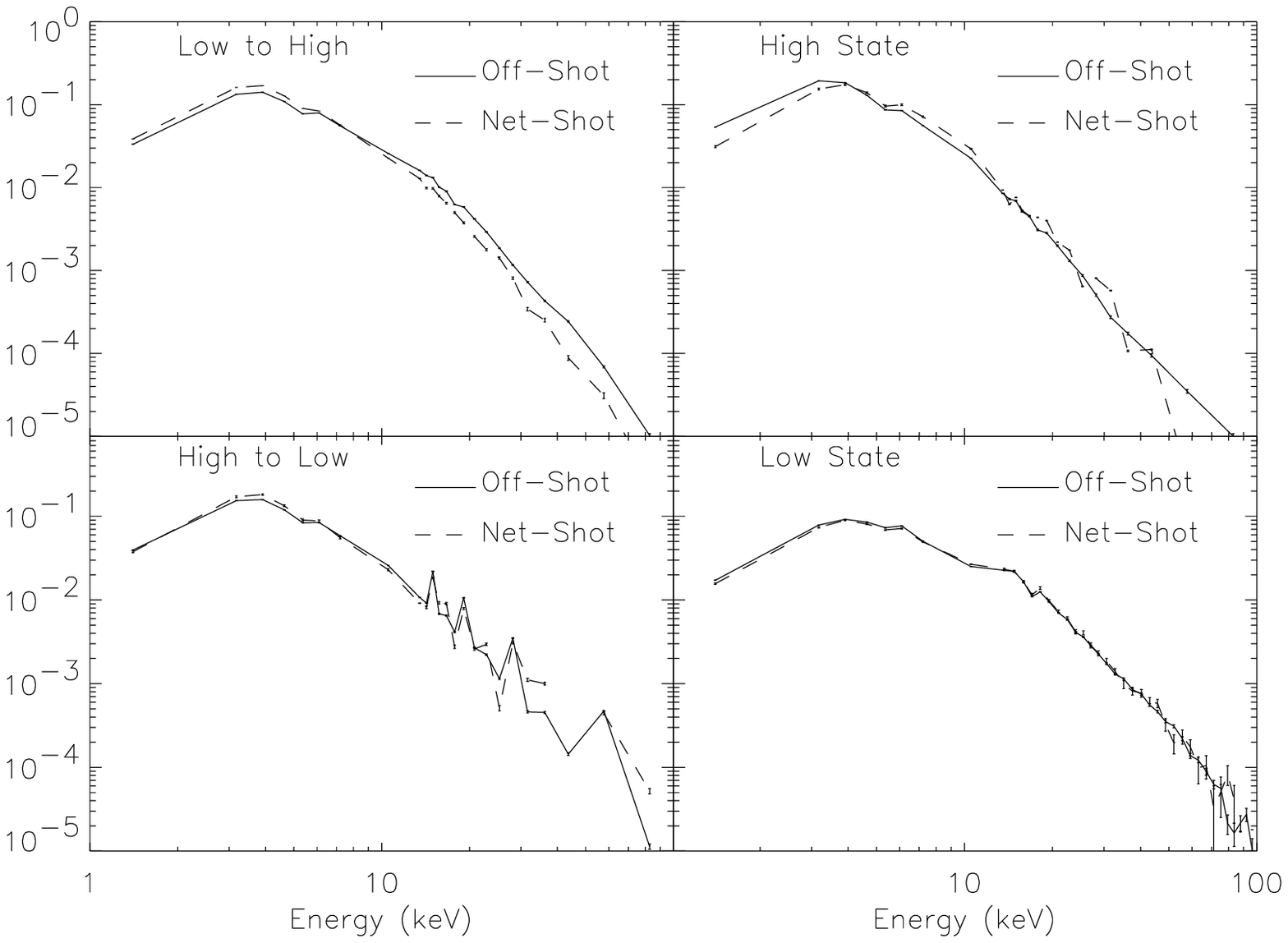,width=15cm,height=15cm,angle=0,%
%bbllx=100,bblly=370,bburx=496,bbury=542
}
\caption{ The net-shot and off-shot spectra with normalized intensity
in the different states.}
\end{figure}
Figure 12 shows the net-shot spectra in the different states for comparison.
It can be seen that the spectra of net-shot 
in the low state is the hardest one. 

\begin{figure}
\hbox{
\epsfig{figure=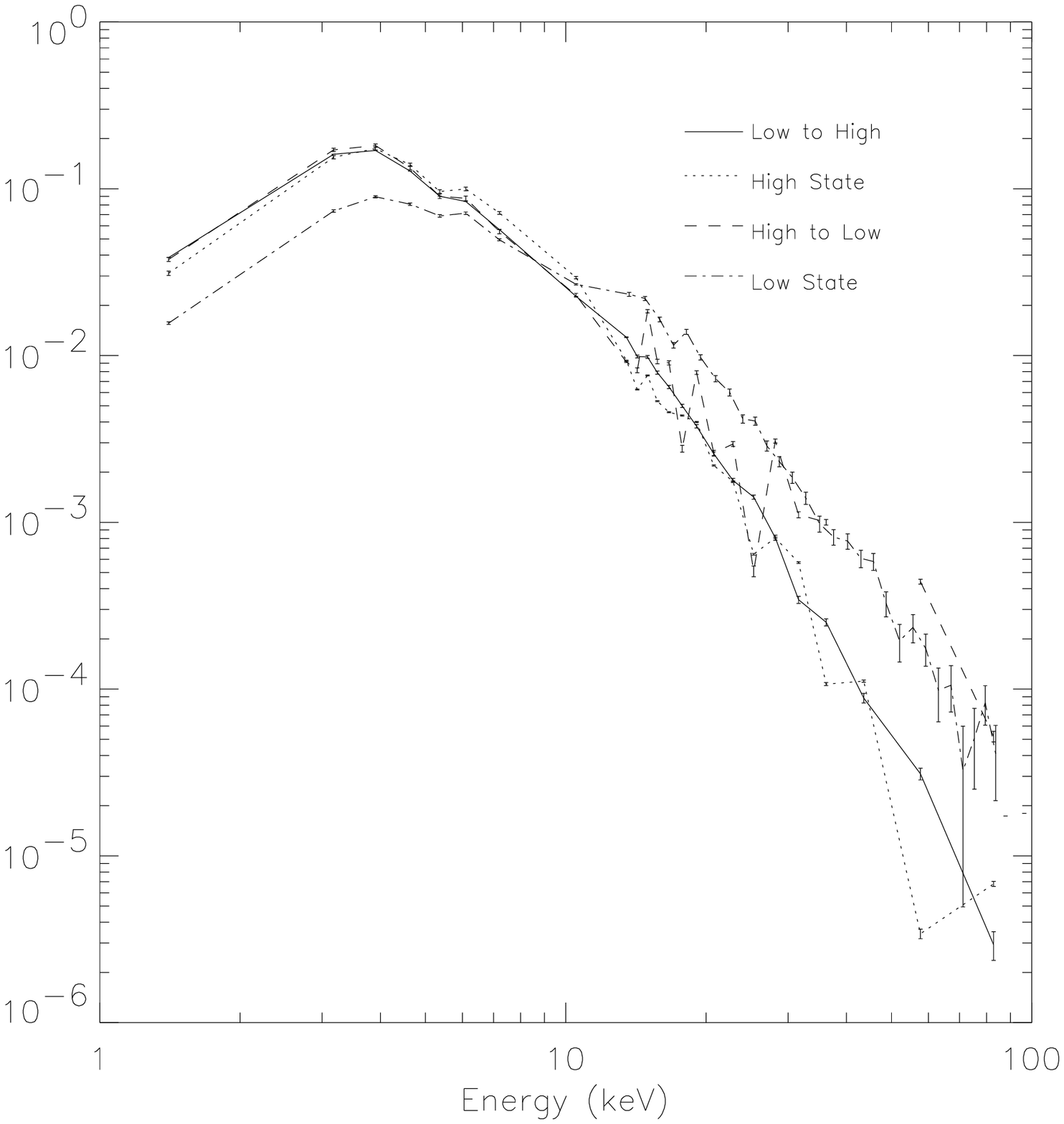,width=10cm,height=6cm,angle=0,%
%bbllx=100,bblly=370,bburx=496,bbury=542
}
\parbox[b]{5cm}{
\caption{ Net-shot spectra with normalized intensity in different states.}
}}
\end{figure}

\section{Discussion}
 The shot durations found from observation of PCA/{\sl RXTE} are 
significantly shorter than that obtained with {\sl Ginga} data by NMK.
The longer time constant of shots found from PCA/{\sl RXTE} observations 
is similar to the shorter time constant obtained by NMK, while 
the shorter time constant obtained by our studies is shorter than the time 
resolution of {\sl Ginga}. The typical interval between the two successive shots 
is about $0.4-0.5$ s. The longer duration found by NMK
could be due to the relatively lower time resolution in the {\sl Ginga} data and the
overlapping effect of shots in their superposing process.
The average shot profile obtained by superposing the simulated data shows
that such methods tend to smooth the structure of individual shot. Although
the profiles of shot are not identical as shown in Figure 3, the shots with
different profiles share some common features, such as the hardness evolution
properties and rapid decay within $\sim$ 20 ms of shot peak, are reflected by
the average shot. 

The shot features vary in different energy bands even in the same state.
The duration of shot in the high
energy band is  shorter than that in the low energy bands, as seen
in Figure 6, and it is more significant in the transition and high states than
that in the low state. It is not consistent with the prediction of the models 
in which X-ray shots are produced by inverse-Compton scattering of flares of 
soft radiation. Because the higher the energy of photons is, the more the 
scatterings for the low energy photons are needed (Sunyaev \& Titarchuk 1980; 
Hua et al. 1997; Bottcher \& Liang 1998). The reirradiation of the high 
energy photons by cool matter on disk could be a explanation 
of that the duration of shot is larger in the low energy bands than that 
in the high energy bands, but it will cause a lose of the near-perfect 
coherence between the low- and high-energy x-ray photons observed in X-ray 
emission of Cyg X-1 (Vaughan \& Nowak 1997). The propagation model 
(Manmoto et al. 1996) assumes that an X-ray shot comes from a disturbance 
added onto the advection-dominated disk at large radii, which propagates 
inward and is reflected at the innermost part of the disk as an outgoing 
shock wave. This model can explain some shot features obtained by NMK. 
In the disturbance propagation model the observed main shot properties, i.e. the shot duration 
is shorter and its peak is later than that in the low energy, could be naturely 
attributed to that the higher temperature the region is, the more inner and 
smaller it is, if the temperature rises gradually from the  outer to inner 
region of the disk. But the propagation model expects the resulting light profile 
to be roughly symmetric with respect to the shot peak, i.e., the rise and decay
parts should be roughly the same, which is not consistent with the asymmetric structure 
obtained from the observation of {\sl RXTE} presented in this paper.

Our results show that some temporal and spectral characteristics 
of X-ray shots of Cyg X-1 are remarkably different in the different states.
The shot duration is the shortest in the high state and the longest in the
low state. The evolution of hardness during a shot in the  high state 
is significantly different from that in the transition and low states. 
It could be a new criterion to distinguish the high state from the other states 
of Cyg X-1 and interesting to apply this criterion to other black hole 
binaries to judge their states. No proposed shot model has predicated  
such a difference in temporal and spectral properties between 
the high state and other states. 
The factor of more effective cooling in the disturbance has been used 
to explain the pre-peak drop of hardness during a shot in the transition 
and low states, but it could not explain the hardness
 during a shot being higher than that of the steady emission 
in the high state. 
It may imply that the environment of the region where the rapid fluctuation 
happens is changed with the spectral state.
The observed hard X-ray emission may mainly be produced by Comptonization
in a hot advection-dominated accretion flow (ADAF) (Esin et al. 1997) or 
an accretion disk corona (ADC) (Dove et el. 1997). The joint region 
between the innermost orbit of the disk and the hot corona could 
be the place where most violent disturbances occur and X-ray shots come from.
In the low or intermediate state, the hot flow or corona can be seen as a
sphere surrounding the black hole horizon with an extending accretion disk
and the joint region is the outer part of the hot corona with a lower temperature
of electrons. On the other hand, in the high state, the ADAF or ADC is restricted to the 
corona above the disk (Esin et al. 1997) and the joint region is then the inner 
part of the corona with a higher temperature. The geometries of hot corona
may explain the difference of hardness evolution in the different states. 
The shot duration may relate to the size of the hot corona. The shot durations 
obtained in the different states implies that the size of corona is smaller 
in the high state than that in the transition states, and it is the largest 
in the low state, which is also consistent with the ADAF or ADC model. 

The time lag of high energy photons relative to low energy ones is a useful probe 
of the dynamics of the hard X-ray production region. 
Although a time delay of X-ray shots in the high energy band 
is, on the average, observed for each state of Cyg X-1 (see Table 2), 
which is consistent with results from the Fourier cross spectral analyses, 
but the features of  the distributions of time lags are complicated 
and difficult to be explained with any simple production model.  
A futher sdudy of the time lag distributions is in the process.   
 
We thank James Lochnner for his help in obtaining and reducing the data, Shuangnan Zhang
for his valuable comments, and Fanjin Lu and 
Yong Chen for their worthful discussions. This research has made use of data 
obtained through the High Energy Astrophysics Science Archive Research 
Center Online Service, provided by the NASA/Goddard Space Flight Center. We also 
thank the anonymous referee for his very useful comments to improve our study.

\section*{References}
\begin{verse}
Bai, T., $\&$ Dennis, B. 1985, ApJ, 292, 699\\
Belloni, T., Mendez, M., Van der Klis, M., Hasinger, G., 
Lewin, W. H. G.,  $\&$Van Paradijs, J. 1996, ApJ, 472, L107\\
Bottcher, M., Liang, E. P. 1998, astro-ph/9802283\\
Bradt, H. V., Rothschild, 
R. E., $\&$ Swank, J. H. 1993, A$\&$A suppl. ser. 97, 355\\
Cui, W., Heindl, W. A., Rothschild, R. E., et al. 1997a, ApJ, 474, L57\\
Cui, W., Zhang, S. N., Focke, W., $\&$ Swank, J. H. 1997b, ApJ, 484, 383\\
Diaconis, P., $\&$ Efron, B. 1983, Scientific American, May, p96\\
Dove, J. B., Wilms, J., Maisack M. M., $\&$ Begelman, M. C. 1997, ApJ, 487, 747\\
Esin, A. A., Narayan, R., Cui, W., et  al. 1997, ApJ\\
Galeev, A. A., Rosner, R.,  $\&$ Vaiana, G. S. 1979, ApJ, 229, 318\\
Hua, X.-M., Kazanas, D., $\&$ Cui, W. 1997 ApJ (submitted), astro-ph/9710184\\
Li, H., \& Fenimore E. E. 1996, ApJ, 469, L115 (LF)\\
Lightman, A. P., $\&$ Eardley, D. M. 1974, ApJ, 187, L1\\
Lochner, J. C., Swank, J. H., $\&$ Szymkowiak, A. E. 1991, ApJ, 375, 295\\
Manmoto, T., 
Takhuchi, M., $\&$ Minhshige, S., et al. 1996, ApJ, 464, L135\\
Miyamoto, S., Kitamoto, S., Mitsuda, K.,  $\&$ Dotani, T. 1988, Nature, 336, 
450\\
Miyamoto, S., $\&$ Kitamoto, S. 1989, Nature, 342, 773\\
Miyamoto, S., Kitamoto, S., Iga, S., et al. 1992, ApJ, 391, L21\\
Morgan, E. H., Remillard, R. A., $\&$ Greiner, J. 1997, ApJ, 482, 993\\
Narayan, R., $\&$ Yi, I. 1995, ApJ, 452, 710\\
Negoro, H., Miyamoto, S., $\&$ Kitamoto, S. 1994, ApJ, 423, L127 (NMK)\\
Nolan, P. L., et al. 1981, ApJ, 246, 494\\
Nowak, M. A., Dove, 
J. B., Vaughan, B. A., Wilms, J., $\&$ Begelman, M. C. 1997, astro-ph/9712106\\
Oda, M. 1977, Space Sci. Rev., 20, 757\\ 
Pudritz, R. E., $\&$ Fahlman, G. G. 1982, MNRAS, 198, 689\\
Rothschild, R. E., Boldt, E. A., Holt, S. S., 
$\&$ Serlemitsos, P. J. 1974, ApJ, 189, L13\\
Sunyaev, R. A., \& Titarchuk, L. G. 1980, A$\&$A, 86, 121\\
Sutherland, P. G., Weisskopf, M. C., $\&$ Kahn, S. M. 1978, ApJ, 219, 1029\\
Shibazaki, N., $\&$ Hoshi, R. 1975, Prog. Theor. Phys., 54, 706\\
Tanaka, Y., $\&$ Lewin, W. H. G., 
1995, in "X-ray Binaries", eds. W. H. G. Lewin, J. van Paradijs, 
E. P. J. van den Heuvel (Cambridge U. Press, Cambridge) p.126\\
Terrel, N. J. 1972, ApJ, 174, L35\\
Van der Klis, M. 1995, in "X-ray Binaries", 
eds. W. H. G. Lewin, J. van Paradijs, E. P. J. van den Heuvel 
(Cambridge U. Press, Cambridge) p.252\\
Vauhgan, B. A., $\&$ Nowak, M. A. 1997 ApJ, 474, L43\\
Zhang, S. N., 
Harmon, B. A., Paciesas, W. S., $\&$ Fishman, G. J. 1996, IAU Circ 6405\\
Zhang, S. N., Mirable, I.
 F., Harmon, B. A., et al. 1997, In: Proe. of The 4th Compton Symp., 
Williamsburg, VA; AIP conf. Proc. 410, p.141\\
Zhang, S. N., Cui, W., Harmon, B. A., et al. 1997,  ApJ, 477, L95\\
Zhang, W. et al. 1995,  ApJ, 449, 930   \\
Zhang, W., Strohmayer, 
T. E., $\&$ Swank, J. H. 1997, ApJ, 482, L167\\
\end{verse}
\newpage
\begin{center}
Table 1. Best fit parameters for the superposed shot in different states \\

\vspace{4mm}
\begin{tabular}{c c c c c c c} 
\cline{1-7} \\ 
 & \multicolumn{2}{c}{2-6 keV} & \multicolumn{2}{c}{6-13 keV} 
& \multicolumn{2}{c}{13-60 keV} \\
 & - & + & - & + & - & + \\ 
\cline{1-7}
\multicolumn{7}{c}{Low~~ to~~ High} \\ 
\cline{1-7}
$A$ (cts/ms) & 2279$\pm$21 & 1533$\pm$104 & 842$\pm$14
& 676$\pm$22 & 422$\pm$14 & 324$\pm$28 \\
$\tau$ (ms) & 45.6$\pm$0.6 & 25.4$\pm$2.1 & 42.6$\pm$1.1
& 33.9$\pm$1.8 & 17.6$\pm$0.8 & 6.7$\pm$1 \\
$B$ (cts/ms) & ... & 666$\pm$107 & ... & 147$\pm$17 
& ... & 150$\pm$15 \\
$\eta$ (ms) & ... & 106.8$\pm$15.7 & ... & 233.3$\pm$58.2
& ... & 66.5$\pm$1 \\          
$C$ (cts/ms) & 3421$\pm$4 & 3360$\pm$10 & 1511$\pm$3 & 1478$\pm$12  
& 583$\pm$1 & 585$\pm$2   \\
reduced $\chi^{2}/450$ & 1.43 & 0.35 & 1.12 & 0.35 & 0.99 & 0.32 \\ 
\cline{1-7}
\multicolumn{7}{c}{High~~~~~ State} \\
\cline{1-7}
$A$ (cts/ms) & 1541$\pm$46 & 1397$\pm$49  & 743$\pm$27 & 567$\pm$44     
& 262$\pm$16 & 292$\pm$29   \\
$\tau$ (ms) & 14.4$\pm$0.8 & 16.1$\pm$1.1 & 14.8$\pm$1.1 & 6.6$\pm$0.7  
& 12.8$\pm$1.6 & 5.2$\pm$0.8 \\
$B$ (cts/ms) & 924$\pm$28 & 778$\pm$39 & 379$\pm$17 & 505$\pm$16    
& 113$\pm$10 & 163$\pm$12   \\
$\eta$ (ms) & 141.0$\pm$9.2 & 125.4$\pm$10.2 & 135.3$\pm$12.1 & 82.0$\pm$3.8 
& 117.3$\pm$17.6 & 55.7$\pm$4.7 \\
$C$ (cts/ms) & 3481$\pm$12 & 3558$\pm$11 & 1078$\pm$7 & 1121$\pm$3 & 
350$\pm$3 & 372$\pm$1 \\
reduced $\chi^{2}$/450 & 1.22 & 1.23 & 1.31 & 1.58 & 1.06 & 1.05 \\     
\cline{1-7}
\multicolumn{7}{c}{High~~ to~~ Low} \\
\cline{1-7}
$ A$ (cts/ms) & 1459$\pm$19 & 282$\pm$90 & 592.2$\pm$1.4 & 257$\pm$49 
& 213$\pm$11 & 224$\pm$23  \\
$\tau$ (ms) & 32.3$\pm$0.6 & 20.2$\pm$4 & 23.6$\pm$0.8 & 37.5$\pm$1.2
& 15.5$\pm$1 & 5.4$\pm$0.9 \\
$B$ (cts/ms) &  ... & 978$\pm$90 & ...& 395$\pm$20 & ... & 72$\pm$9 \\
$\eta$ (ms) & ... & 42.1$\pm$1.8 & ... & 42.9$\pm$2.3 & ... & 64.1$\pm$10 \\
$C$ (cts/ms) & 1897$\pm$2 & 1896$\pm$3 & 752$\pm$1 & 759$\pm$2   
& 280$\pm$1 & 284$\pm$1   \\
reduced $\chi^{2}$/450 & 1.22 & 0.45 & 1.17 & 0.35 & 1.28 & 0.36 \\
\cline{1-7} \\
& \multicolumn{2}{c}{2-5 keV} & \multicolumn{2}{c}{5-13 keV}
& \multicolumn{2}{c}{13-60 keV} \\
 & - & + & - & + & - & + \\
\cline{1-7}
\multicolumn{7}{c}{Low~~~~~  State}  \\
\cline{1-7} 
$A$ (cts/ms) & 698$\pm$50 & 631$\pm$57 & 1317$\pm$90 & 893$\pm$82     
& 558$\pm$37 & 585$\pm$50   \\
$\tau$ (ms) & 15.3$\pm$1.9 & 5.5$\pm$0.8 & 21.4$\pm$1.8 & 4.5$\pm$0.7 
& 14.6 $\pm$1.7 & 5.0$\pm$0.7 \\
$B$ (cts/ms) & 879$\pm$51 & 1024$\pm$26 & 1081$\pm$95 & 1663$\pm$28    
& 465$\pm$36 & 633$\pm$18    \\
$\eta$ (ms) & 81.0$\pm$4.3 & 59.1$\pm$1.7 & 91.1$\pm$6.6 & 61.5$\pm$1.3 
& 83.4$\pm$6.4 & 64.6$\pm$2.3 \\
$C$ (cts/ms) & 1712$\pm$4 & 1691$\pm$2 & 2662$\pm$7 & 2664$\pm$3   
& 1138$\pm$3 & 1136$\pm$2    \\
reduced $\chi^{2}$/450 & 0.36 & 0.37 & 0.35 & 0.35 & 0.34 & 0.43 \\
\cline{1-7}
\end{tabular}
\end{center}
\newpage
{\small
\centerline{{Table 2. Average hard X-ray lags of shots in different states.}}}
\vspace{1mm}
\begin{center}
\begin{tabular}{c c}
\hline \\
             State & Time Lag (ms) \\
\\ \hline \\
             Low to High     & 9.9$\pm$0.9  \\

             High            & 3.9$\pm$1.0  \\

             High to Low     & 9.1$\pm$1.5  \\
      
             Low             & 1.8$\pm$0.7  \\
\\ \hline \\

\end{tabular}
\end{center}

\end{document}